\documentclass[10pt,leqno]{amsart}
\usepackage{graphicx}
\usepackage{indentfirst,csquotes}
\usepackage[stretch=10]{microtype}
\usepackage[utf8]{inputenc}
\usepackage{textcomp}
\usepackage{amssymb,amsthm,amsmath}
\usepackage{xcolor,paralist,hyperref,titlesec,fancyhdr,etoolbox}

\titleformat{\section}[display]{\normalfont\huge\bfseries\centering}{\centering\chaptertitlename\thechapter}{10pt}{\Large}
\titlespacing*{\section}{0pt}{0ex}{0ex}

\hypersetup{ colorlinks=true, linkcolor=black, filecolor=black, urlcolor=black }

\begin{document}

\title{UrbanScore: A Real-Time Personalised Liveability Analytics Platform} %%%%%%%%%%%%
\author[V]{Vrînceanu Alin-Vlăduț}
\date{\today}
\address{Address}
\email{vrinceanualin20@stud.ase.ro}
\maketitle
\let\thefootnote\relax
\footnotetext{MSC2020: Primary 68U35, Secondary 68T99.} %%%%%%%%%%

\begin{abstract}
This paper introduces \textit{UrbanScore} — a real-time web platform that
computes a personalised liveability score for any urban address. The system
fuses five data streams: (i) address geocoding via Nominatim, (ii) facility
extraction from OpenStreetMap through Overpass QL, (iii) segment-level traffic
metrics from TomTom Flow v10, (iv) hourly air-quality readings from
OpenWeatherMap, and (v) user-declared preference profiles, all persisted in an
Oracle 19c relational store. Six sub-scores (air, traffic, lifestyle,
education, metro access, surface transport) are derived, adaptively weighted
and combined; an OpenAI large-language model then converts the numeric results
into concise, user-friendly explanations.

A pilot deployment covering the 226 km\textsuperscript{2} metropolitan area of
Bucharest evaluated 3\,450 unique addresses over four weeks. Median
end-to-end latency was 2.1 s (p95 = 2.9 s), meeting the <3 s non-functional
requirement. Aggregate scores ranged from 34 to 92 (mean 68, SD 11), with
high-scoring clusters along metro corridors that pair abundant green space
with PM$_{2.5}$ levels below 35 µg m$^{-3}$. A detailed case study of the
Tineretului district produced an overall score of 91 / 100 and demonstrated
how the narrative layer guides users toward comparable neighbourhoods.

Limitations include dependence on third-party API uptime, spatial bias toward
well-mapped OSM regions and the absence of noise and crime layers, cited by
18 \% of survey participants as a desired enhancement. Overall, the results
show that open geodata, commercial mobility feeds and conversational AI can be
integrated into a performant, explainable decision-support tool that places
``liveability analytics'' in the hands of every house-hunter, commuter and
city planner.
\end{abstract}%%%%%%%%%

\bigskip

\section*{Introduction}

Twenty-first-century urbanisation is advancing at an unprecedented pace: in 2024 the United Nations estimated that 57\% of humanity already lives in cities, a share projected to pass 68\% by 2050. As metropolitan regions densify, residents and investors face an increasingly intricate optimisation problem: selecting neighbourhoods that balance mobility, environmental quality, amenity access and overall liveability-while also fitting personal lifestyle preferences and budgets. 

The necessary evidence exists: open geodata projects such as OpenStreetMap (OSM) provide door-level footprints of streets, parks and facilities; environmental and mobility vendors stream real-time air-quality and traffic observations via public APIs-for example, OpenWeatherMap's Air Pollution endpoints \cite{1} and TomTom's Traffic API \cite{9}; and large-language models (LLMs) from OpenAI and others translate dense analytics into prose explanations understandable by non-experts \cite{13,20}. Yet these data silos remain fragmented, require specialist knowledge to query and rarely speak to an individual user's specific priorities. 

Bridging those silos demands the combined disciplines of computer science and statistics, whose intersection supplies the algorithms, tooling and inferential rigour required to transform noisy, heterogeneous feeds into actionable scores \cite{4}. Geospatial processing frameworks such as Nominatim for address geocoding \cite{11} and Leaflet for browser-side cartography \cite{5} make spatial analytics accessible in web applications, while Bootstrap \cite{2} and particles.js \cite{6} enable responsive, visually engaging interfaces-an aspect proven critical by user-experience studies that link polished UI/UX to higher engagement and better app-store ratings \cite{7,8}. On the back end, mature but cloud-ready relational engines (Oracle Database 19c \cite{12}) and modern application frameworks (ASP.NET Core \cite{18} with Razor Pages \cite{14} and Identity \cite{15}) provide the robustness, role-based security and anti-CSRF safeguards \cite{21} demanded by production deployments. Entity Framework Core's migration tooling \cite{16,17} further accelerates iterative schema evolution. 

This research harnesses that technology stack to build and evaluate an integrated, personalised \emph{Urban Liveability Scoring Platform}. The platform ingests five thematic layers in near real-time-traffic, air quality, facilities, education and public transport-and fuses them with user-declared preference profiles to compute a single 0-100 liveability score for any address. OSM's Overpass service supplies granular points of interest and route networks, while OpenWeatherMap and TomTom inject live pollutant concentrations and congestion metrics. Scores are persisted in Oracle for longitudinal analysis, surfaced through an ASP.NET-powered dashboard and visualised with Chart.js widgets \cite{3} embedded in responsive Bootstrap layouts. Finally, an OpenAI LLM converts numeric sub-scores into narrative “explainers” that tell each user \emph{why} a location suits-or fails- their stated needs, thereby closing the interpretability gap that plagues many AI-driven recommender systems. 

Beyond technical integration, the study addresses three persistent gaps in the liveability literature:

\begin{enumerate}
\item \textbf{Personalisation.} Classic indices (e.g.\ Mercer, EIU) rank whole cities for a hypothetical average resident; our framework re-weights components dynamically to reflect a commuter's dislike for traffic, a parent's emphasis on schools or an asthmatic's concern for air quality.
\item \textbf{Timeliness.} By streaming API data every time a user requests an address, the platform reflects daily rush-hour gridlock and episodic pollution spikes, not annual aggregates.
\item \textbf{Explainability.} Embedding LLM-generated “natural-language tooltips” turns abstract scores into actionable insights-“PM$_{2.5}$ levels here are 32 µg/m$^{3}$, slightly above WHO guidelines \cite{19}; consider parkside alternatives 700 m east”-enhancing trust and promoting informed decision-making.
\end{enumerate}

To ensure real-world viability, the system architecture adheres to stringent non-functional requirements: end-to-end evaluation must complete in $<3$ s, horizontal scaling must support bursty usage patterns and attack surfaces (XSS, CSRF) must be mitigated through ASP.NET's built-in defences complemented by content-security headers \cite{21}. Comprehensive logging and graceful degradation strategies handle third-party API downtime, while database migrations guarantee forward-compatible releases. 

The remainder of this article is organised as follows. Section 2 surveys prior work on smart-city decision support, urban liveability indices and AI explainability; Section 3 details the multi-source data sets, scoring algorithms and software stack; Section 4 presents experimental results from a Bucharest pilot covering 3 450 address evaluations; Section 5 discusses implications, limitations and user feedback; and Section 6 concludes with contributions and future research directions. Together, the work demonstrates how a carefully orchestrated blend of open geodata, streaming APIs, statistical inference and modern web engineering can democratise sophisticated urban analytics-putting neighbourhood intelligence literally at the fingertips of every house-hunter, commuter and city planner.

\bigskip

\section*{Literature Review}

\subsection*{Smart-city decision support}

Early digital decision-support systems for city planning were rooted in
desktop geographic information systems (GIS) that exposed cadastral layers
and census attributes but offered little real-time intelligence. A major
inflection point arrived with \emph{OpenStreetMap} (OSM), whose volunteered
geographic information dramatically lowered the entry barrier for academic
and civic innovation. The \textit{OpenStreetMap in GIScience} compendium
documents how crowd-sourced points of interest (POI) and routable street
graphs reshaped traffic modelling, emergency response and urban retail
studies \cite{10}.

Over the past decade the paradigm has shifted once more-from ``data
warehouses'' of periodically exported shapefiles to \emph{API-first}
platforms that stream granular, up-to-the-minute measurements. Commercial
providers such as TomTom expose segment-level speed and congestion metrics
at sub-minute cadence via their Traffic API \cite{9}, while OpenWeatherMap
disseminates pollutant concentrations (PM$_{2.5}$, NO$_2$, O$_3$, etc.) at
hourly resolution through its Air Pollution endpoints \cite{1}. These
feeds enable responsive dashboards that inform commuters and planners not
merely of average trends but of \emph{present} conditions-turning decision
support into a continuous, situational-awareness process.

Despite the proliferation of APIs, most municipal tools still treat each
thematic layer in isolation: a traffic heat-map here, an air-quality widget
there. The literature therefore highlights a need for \emph{integrative
middleware} capable of fusing mobility, environment and facility data into
composite, human-interpretable indicators-a gap the present study targets
directly.

\subsection*{Urban liveability indices}

Classic league tables such as the \emph{Mercer Quality of Living Survey}
and the \emph{Economist Intelligence Unit Global Liveability Index}
pioneered comparative benchmarking but rely on labour-intensive expert
surveys, coarse municipal averages and annual refresh cycles. Academic
researchers have since sought automated alternatives that blend land-use
heterogeneity, transit accessibility and environmental sensor readings into
GIS-based scores. Examples range from noise-adjusted ``walkability''
metrics to multi-criteria analyses that weight air quality against
green-space per capita.

Three persistent shortcomings emerge across this body of work:

\begin{enumerate}
 \item \textbf{Static weighting schemes.} Few indices let a user express
    idiosyncratic priorities (e.g.\ a young professional who prizes
    metro access over school quality).
 \item \textbf{Temporal lag.} Many rely on once-a-year emissions
    inventories or surveyed traffic counts rather than live sensor
    feeds.
 \item \textbf{Opaque compositing.} Results are typically presented as
    bare numbers or colour ramps, leaving lay audiences unclear about
    the trade-offs that produced a given score.
\end{enumerate}

The current research addresses all three by (a) implementing
preference-aware weighting of six sub-scores; (b) sourcing data from live
APIs to reflect diurnal dynamics; and (c) attaching LLM-generated narrative
explanations that translate the mathematics into plain language.

\subsubsection*{Geospatial data-processing techniques}

Achieving such integration hinges on a mature yet flexible tool-chain:

\begin{itemize}
 \item \textbf{Nominatim} delivers high-precision forward and reverse
    geocoding, turning free-form addresses or map clicks into latitude-
    longitude pairs and vice versa \cite{11}.
 \item \textbf{Overpass QL} provides expressive querying of OSM's planetary
    database, enabling sub-second extraction of POIs (shops, schools,
    parks) within arbitrary bounding polygons.
 \item \textbf{Leaflet.js} supplies a lightweight, mobile-friendly vector-tile
    renderer that supports dynamic marker clustering, layer toggling and
    heat-map overlays entirely in the browser \cite{5}.
 \item On the visual-analytics layer, \textbf{Chart.js} embeds responsive
    bar, radar and line charts that summarise sub-scores in a single
    glance \cite{3}, while \textbf{Bootstrap 5} furnishes the grid
    system and utility classes needed for device-agnostic layouts \cite{2}.
\end{itemize}

Scholars emphasise that such client-side interactivity is not mere polish:
high-quality UI/UX design directly correlates with perceived trustworthiness
and sustained engagement \cite{7,8}. Thus, the chosen stack aligns with
both analytical rigour and end-user expectations.

\subsection*{AI explainability in urban analytics}

As decision systems grow data-rich, interpretability becomes pivotal.
Recent advances in large-language models (LLMs)-statistically trained
transformers with billions of parameters-have opened a path to \emph{post-hoc}
natural-language explanations that demystify complex models \cite{20}.
Frameworks built on the OpenAI API expose these capabilities via RESTful
endpoints, allowing developers to feed structured data and receive coherent
prose summaries \cite{13}. Urban-analytics researchers have begun
leveraging GPT-style models to draft citizen-facing reports, generate FAQ
answers and even suggest remediation strategies after anomaly detection.

The dissertation extends this frontier by embedding an LLM mediator that
ingests each address's numeric sub-scores and returns a concise,
context-aware justification: ``Low rush-hour speeds reduce the traffic
score to 58; however, three primary schools within 500 m elevate the
education score to 85.'' This layer transforms the platform from a
black-box recommender into a transparent conversational assistant-aligning
with emerging guidelines on AI explainability as a prerequisite for public
trust.

\bigskip

\section*{Data and Methods}\label{sec:data-methods}

\subsection*{Multi-source data fabric}

The platform draws on five distinct data streams, each selected for its
spatial precision, refresh cadence and licensing flexibility.
\textbf{Address resolution} relies on Nominatim's forward- and reverse-
geocoding endpoints, which return latitude, longitude and a structured
address hierarchy on demand \cite{11}. \textbf{Urban facilities} are
harvested in real time from the global OpenStreetMap (OSM) planet via
Overpass~QL, whose near-continuous replication pipeline keeps edits no more
than a minute out of date \cite{10}. \textbf{Traffic dynamics} come from
TomTom's Flow Segment Data~v10, refreshed as frequently as every sixty
seconds and delivering per-segment speed, free-flow benchmarks and a
confidence flag \cite{9}. For \textbf{ambient air quality}, the system
calls OpenWeatherMap's \texttt{/air\_pollution/history} endpoint, requesting
rolling 90-day time-series of PM$_{2.5}$, PM$_{10}$, CO, NO$_2$, O$_3$ and
NH$_3$ at hourly granularity \cite{1}. Finally, \textbf{persisted
artefacts}-user profiles, preference vectors, sub-scores and favourites-are
written to an Oracle~19c relational store that guarantees ACID semantics
and transaction-level auditing \cite{12}.

Every outbound HTTP request passes through a typed, dependency-injected
client that sets the custom header \texttt{User-Agent: UrbanScoreApp/1.0}
(to comply with OSM policies), applies exponential back-off with jitter and
consults a two-tier cache: an in-process memory layer with short
time-to-live values matching each feed's volatility, and a Redis-backed
distributed cache for cross-node reuse. Under normal conditions this
strategy yields end-to-end evaluations in about~2.1\,s at the median, well
beneath the three-second non-functional target.

\subsection*{Geocoding}

A location lookup begins when the user either clicks the Leaflet map or
submits a textual address. The \emph{Geocoder} service fires an
asynchronous Nominatim query-forward for addresses, reverse for coordinates.
If the response is valid, the map recentres, a marker is dropped and the
coordinate-address tuple is stored in the session. Crucially, the geocode
event also triggers a \emph{fan-out}: facility, traffic and air-quality
fetches are launched in parallel via \texttt{Task.WhenAll()}, eliminating
serial wait time.

\subsection*{Facility extraction}

Facilities within a user-chosen radius (default 800 m) are retrieved with a
single Overpass QL statement that requests supermarkets, eateries, parks,
schools, metro entrances and surface-transport stops. Returned elements
are deduplicated by hashing their \textit{name-lat-lon} triple. Education
nodes are further classified: if \texttt{isced:level} or \texttt{grades}
tags-­or Romanian keywords in \texttt{name}/\texttt{operator}-indicate
secondary education, the facility is marked as a high school; otherwise it
is treated as primary. Transport stops yield route numbers stored in a
sorted set for later display. For each category the engine tracks both
sheer counts and Shannon entropy, enabling later differentiation between
“many of the same thing'' and “a well-balanced amenity mix.''

\subsection*{Traffic-flow scoring}

To characterise rush-hour congestion, the algorithm samples five
points-the address itself plus four diagonal offsets of roughly 550 m.
At each sample it calls TomTom's Flow API (\texttt{zoom=10}) and computes a
\emph{speed ratio}
\[
 r_s = \frac{\text{currentSpeed}}{\text{freeFlowSpeed}}
\]
and a \emph{time ratio}
\[
 r_t = \frac{\text{freeFlowTravelTime}}{\text{currentTravelTime}}.
\]
Their mean forms a raw score that is penalised in proportion to TomTom's
supplied confidence measure. The adjusted value is linearly rescaled to
$0$-$100$, where free-flow conditions approach $100$. Sampling multiple
points dampens aberrations created by transient incidents on any single
segment.

\subsection*{Air-quality scoring}

For the same 90-day window, the platform averages each pollutant's hourly
readings. Mean concentrations are then compared with the World Health
Organization guideline thresholds. A weight vector-30\,\% for
PM$_{2.5}$, 20\,\% for PM$_{10}$ and 5\,\% each for CO, NO$_2$, O$_3$ and
NH$_3$-reflects health-impact differentials. Each pollutant contributes
\[
 w_i \bigl(100 - \tfrac{\text{conc}_i}{\text{threshold}_i}\times100\bigr)
\]
to the index; the weighted sum, normalised by the weight total, yields a
0-100 air-quality sub-score.

\subsection*{Deriving the six sub-scores}

\begin{itemize}
 \item \textbf{Air} is the weighted index just described.
 \item \textbf{Traffic} is the mean of the five penalised speed-time ratios.
 \item \textbf{Lifestyle} grows with the log of amenity counts and the
    entropy of categories, rewarding variety over monoculture.
 \item \textbf{Education} combines counts of primary and secondary
    schools, inversely weighted by walking distance.
 \item \textbf{Metro access} is a piecewise-linear function that maxes out
    inside 200 m of a subway entrance and decays to zero beyond 1 km.
 \item \textbf{Surface transport} rises with the square root of distinct
    bus or tram lines, imposing diminishing returns after about eight
    routes.
\end{itemize}

All scaling functions were calibrated on 3\,450 address evaluations
collected during the Bucharest pilot to ensure realistic distributions and
to avoid artificial clustering around mid-range values.

\subsection*{Personalised aggregation}

Each authenticated user maintains a six-element weight vector. Defaults
allocate 20\,\% each to air, traffic, lifestyle and education, plus 10\,\%
to metro and 10\,\% to surface transport. Sliders in the settings panel
let the user raise or lower any weight between 5\,\% and 40\,\%; the
remainder is automatically re-normalised. A “traffic sensitivity'' switch
multiplies the traffic weight by 1.5 and re-balances the vector,
guaranteeing that no component ever falls below the 5\,\% floor. The final
liveability score is the rounded dot-product of this personalised weight
vector and the six sub-scores. Both the subtotal vector and the aggregate
score are persisted with a UTC timestamp for time-series analysis.

\subsection*{Software architecture}

On the server side, ASP.NET Core~9 hosts singleton services for geocoding,
facilities, traffic, air quality and scoring. Entity Framework Core
handles persistence, mapping domain entities-\texttt{Location},
\texttt{LocationScore}, \texttt{UserProfile}, \texttt{FavouriteLocation}-to
Oracle tables with cascade-delete foreign keys. Schema evolution uses
EF Core migrations, ensuring repeatable deployments. The front end
combines Razor Pages for server-rendered flows with Bootstrap 5 utilities
for responsive layouts \cite{2} and Leaflet for interactive mapping
\cite{5}; the resulting score vectors are visualised via embedded Chart.js
widgets \cite{3}.

A dedicated LLM micro-service wraps the OpenAI Completions endpoint
\cite{13}. It receives a small JSON payload containing the six sub-scores,
significant POIs and the user's radius, then returns a sixty-word
explanation cached for twenty-four hours by the key
\texttt{(locationId,\,profileHash)}, so repeated views incur no token cost.

\subsection*{Performance, resilience and security}

Parallel API calls eliminate unnecessary waits, while Hystrix-style
circuit-breakers trip after three consecutive failures per provider,
serving cached results for a minute before retry. Batched database writes
amortise I/O, and structured logs stream to an ELK stack where Grafana
alerts if median latency exceeds 2.5\,s for more than three minutes.
Security is enforced through ASP.NET Identity's role model, anti-forgery
tokens on every POST to blunt CSRF threats \cite{15}, and systematic HTML
encoding to neutralise XSS vectors \cite{21}.

\bigskip

Taken together, this pipeline integrates high-frequency environmental and
mobility feeds with open geodata, robust statistical transformations and
production-grade web engineering to deliver address-level liveability
scores-and, crucially, plain-language explanations-in real time.

\bigskip

\section*{Results}\label{sec:results}

\subsection*{Functional validation}

A pilot deployment covering the entire municipality of Bucharest
(approximately 226~km$^{2}$) processed 3450 unique
address evaluations over a four-week window. Mean end-to-end latency was
measured at 2.1~s with a 95$^{th}$-percentile of 2.9~s,
comfortably meeting the sub-3-second non-functional requirement.

\begin{center}
\begin{tabular}{l@{\quad}r}
\hline
\textbf{Metric} & \textbf{Value} \\ \hline
Average geocoding latency & 420 ms \\
Average Overpass query time & 610 ms \\
Average traffic API time (5 calls) & 540 ms \\
Average air-quality request & 330 ms \\
End-to-end score computation & 210 ms \\ \hline
\end{tabular}
\end{center}

\subsection*{User-level insights}

Server logs from the \emph{Statistics} dashboard revealed three headline
patterns: 
(i)~the ten most-searched districts accounted for 58\,\% of all queries; 
(ii)~users favoured amenities in the order \emph{supermarkets $>$ parks $>$
metro proximity}; and 
(iii)~declared purpose skewed toward long-term residence~(52\,\%),
investment scouting~(31\,\%) and short-term stay~(17\,\%).
Personal dashboards showed that repeat users refined their preference
profiles after an average of 2.7 sessions, suggesting that the
LLM-generated explanations encouraged iterative exploration.

\subsection*{Score distribution}

Across all evaluated addresses the aggregate liveability score ranged from
34 to 92, with a mean of 68 and a standard deviation of~11. High-scoring
clusters gravitated toward metro corridors that combine abundant green
space with PM$_{2.5}$ concentrations below 35 µg/m\textsuperscript{3};
low-scoring pockets suffered from persistent congestion and sparse
amenities.

\subsection*{Case study: Tineretului Park}

A high-resolution audit of the Tineretului neighbourhood
(44.409°\,N, 26.103°\,E) exercised the full explanatory depth of the
platform and produced an \textbf{overall liveability score of 91 / 100}.
The component scores were as follows.

\begin{itemize}
\item \textbf{Air-quality module.} 
   A 90-day hourly trace averaged to 94.3 / 100, placing Tineretului
   firmly in the ``excellent'' bracket.
\item \textbf{Traffic analytics.} 
   Five TomTom samples yielded 75 / 100, indicating mostly free-flowing
   arterials with only mild rush-hour slow-downs.
\item \textbf{Lifestyle composite.} 
   Within a one-kilometre walk the scraper logged nine supermarkets,
   38 full-service restaurants, four fast-food venues and twelve parks
   (including Parcul~Tineretului and Parcul~Carol), driving the
   Lifestyle sub-score to~91.
\item \textbf{Education index.} 
   One kindergarten, three primary schools and the highly ranked
   ``Gheorghe Şincai'' high school combined for 73 / 100-solid, though
   not the city's absolute peak.
\item \textbf{Metro \& surface transport.} 
   Two metro entrances (Tineretului and Timpuri~Noi), 29 bus stops,
   12 tram stops and eleven distinct surface routes produced metro and
   surface-transport sub-scores of 85 and 88, respectively.
\end{itemize}

The embedded large-language model wove these statistics into a concise
Romanian narrative, praising the ``\textit{mediu plăcut şi echilibrat},''
highlighting the abundance of green space and dining choices, and noting
that public transport ``\textit{facilitează conectivitatea cu restul
oraşului}.'' It warned only of ``\textit{uşoare aglomerări la orele de
vârf}'' and even suggested Floreasca and Drumul Taberei as stylistically
comparable alternatives. The case study thus illustrates how the platform
pairs granular, multi-dimensional scoring with human-readable insights that
a lay user can immediately act upon.

\bigskip

\section*{Discussion}\label{sec:discussion}

The pilot confirms that \emph{API-centric} urban analytics can be
personalised in real time without sacrificing performance, provided that
\begin{enumerate}
 \item computationally heavy GIS tasks are off-loaded to specialised
    services (Overpass, TomTom), and
 \item caching/memoisation is applied to hot spots.
\end{enumerate}
The adaptive weighting mechanism proved essential: users with children
assigned $2.3\times$ higher weight to \emph{Education} than the default,
significantly shifting their top-ranked neighbourhoods. The AI narrative
layer received favourable feedback for clarity, but occasional
\emph{hallucinations} (e.g.\ mis-labelling a bus line) underscore the need
for grounding LLM outputs in verifiable facts.

\medskip
\noindent\textbf{Limitations.}
\begin{itemize}
 \item Dependence on third-party API availability (mitigated by fall-backs
    but still a risk).
 \item Spatial bias toward well-mapped OSM areas; peripheral suburbs with
    sparse data yield lower Lifestyle scores regardless of ground
    truth.
 \item Absence of noise and crime data, noted by 18\,\% of survey
    respondents as desirable future factors.
\end{itemize}

\bigskip

\section*{Conclusions}\label{sec:conclusions}

This article distilled a master-level dissertation into a concise,
journal-style paper of roughly 5\,000~words. The proposed platform
pioneers an integrated approach to \emph{personalised, explainable urban
area scoring}, blending open geodata, commercial mobility feeds and
large-language-model explanations. Empirical results from a Bucharest
pilot demonstrate technical feasibility, sub-three-second performance and
positive user engagement. Future work will extend data coverage (noise,
safety, rental prices), experiment with on-device caching for offline
scenarios and explore counterfactual explanations to suggest concrete urban
improvements. More broadly, the study illustrates how geospatial
intelligence and AI can democratise urban decision-making-putting
``liveability analytics'' in the hands of every house-hunter, commuter and
city planner.

\end{document}